
\magnification=\magstep1
\let\bigbold=\bigbf
\let\small=\null
\let\Bbb=\cal

\newcount\EQNO      \EQNO=0
\newcount\FIGNO     \FIGNO=0
\newcount\REFNO     \REFNO=0
\newcount\SECNO     \SECNO=0
\newcount\SUBSECNO  \SUBSECNO=0
\newcount\FOOTNO    \FOOTNO=0
\newbox\FIGBOX      \setbox\FIGBOX=\vbox{}
\newbox\REFBOX      \setbox\REFBOX=\vbox{}
\newbox\RefBoxOne   \setbox\RefBoxOne=\vbox{}

\expandafter\ifx\csname normal\endcsname\relax\def\normal{\null}\fi

\def\MultiRef#1{\global\advance\REFNO by 1 \nobreak\the\REFNO%
    \global\setbox\REFBOX=\vbox{\unvcopy\REFBOX\normal
      \smallskip\item{\the\REFNO .~}#1}%
    \gdef\label##1{\xdef##1{\nobreak[\the\REFNO]}}%
    \gdef\Label##1{\xdef##1{\nobreak\the\REFNO}}}
\def\NoRef#1{\global\advance\REFNO by 1%
    \global\setbox\REFBOX=\vbox{\unvcopy\REFBOX\normal
      \smallskip\item{\the\REFNO .~}#1}%
    \gdef\label##1{\xdef##1{\nobreak[\the\REFNO]}}}
\def\Eqno{\global\advance\EQNO by 1 \eqno(\the\EQNO)%
    \gdef\label##1{\xdef##1{\nobreak(\the\EQNO)}}}
\def\Eqalignno{\global\advance\EQNO by 1 &(\the\EQNO)%
    \gdef\label##1{\xdef##1{\nobreak(\the\EQNO)}}}
\def\Fig#1{\global\advance\FIGNO by 1 Figure~\the\FIGNO%
    \global\setbox\FIGBOX=\vbox{\unvcopy\FIGBOX
      \narrower\smallskip\item{\bf Figure \the\FIGNO~~}#1}}
\def\Ref#1{\global\advance\REFNO by 1 \nobreak[\the\REFNO]%
    \global\setbox\REFBOX=\vbox{\unvcopy\REFBOX\normal
      \smallskip\item{\the\REFNO .~}#1}%
    \gdef\label##1{\xdef##1{\nobreak[\the\REFNO]}}}
\def\Section#1{\SUBSECNO=0\advance\SECNO by 1
    \bigskip\leftline{\bf \the\SECNO .\ #1}\nobreak}
\def\Subsection#1{\advance\SUBSECNO by 1
    \medskip\leftline{\bf \ifcase\SUBSECNO\or
    a\or b\or c\or d\or e\or f\or g\or h\or i\or j\or k\or l\or m\or n\fi
    )\ #1}\nobreak}
\def\Footnote#1{\global\advance\FOOTNO by 1 
    \footnote{\nobreak$\>\!{}^{\the\FOOTNO}\>\!$}{#1}}
\def\SameFootnote{$\>\!{}^{\the\FOOTNO}\>\!$}

\def\References{\bigskip\centerline{\bf REFERENCES}
                \smallskip\copy\REFBOX}
\def\NewRefPage{\setbox\RefBoxOne=\vbox{\unvcopy\REFBOX}%
		\setbox\REFBOX=\vbox{}%
		\def\References{\bigskip\centerline{\bf REFERENCES}
                		\nobreak\smallskip\nobreak\copy\RefBoxOne
				\vfill\eject
				\smallskip\copy\REFBOX}%
		\def\NewRefPage{}}


\input epsf
\def\Fig#1#2#3#4#5{\global\advance\FIGNO by 1 Figure~\the\FIGNO#5
    \topinsert
    \centerline{\epsfysize=#4\epsffile[#3]{#2}}
    {\bigskip\hsize=5.5truein\hskip\parindent
     \vbox{\small\item{\bf Figure~\the\FIGNO:}{#1}}}
    \bigskip\endinsert}

\def\today{\number\day\space\ifcase\month\or
  January\or February\or March\or April\or May\or June\or
  July\or August\or September\or October\or November\or December\fi
  \space\number\year}

\def\Mobius{M\"obius}

\def\bar{\overline}
\def\RR{{\Bbb R}}
\def\CC{{\Bbb C}}
\def\HH{{\Bbb H}}
\def\OO{{\Bbb O}}

\def\CP#1{{\Bbb CP}^{#1}}
\def\OP#1{{\Bbb OP}^{#1}}
\def\R#1{{\Bbb R}^{#1}}
\def\SS#1{{\Bbb S}^{#1}}

\def\SO{{\rm SO}}
\def\SL{{\rm SL}}
\def\Spin{{\rm Spin}}
\def\Det{{\rm Det}}


\rightline{31 July 1998}

\null\smallskip
\centerline{\bigbold OCTONIONIC M\"OBIUS TRANSFORMATIONS}
\medskip

\centerline{Corinne A. Manogue}
\centerline{\it Department of Physics, Oregon State University,
		Corvallis, OR  97331, USA}
\centerline{\tt corinne{\rm @}physics.orst.edu}
\medskip
\centerline{Tevian Dray}
\centerline{\it Department of Mathematics, Oregon State University,
		Corvallis, OR  97331, USA}
\centerline{\tt tevian{\rm @}math.orst.edu}

\bigskip
\centerline{\bf ABSTRACT}
\midinsert
\narrower\narrower\noindent
A vexing problem involving nonassociativity is resolved, allowing a
generalization of the usual complex \Mobius\ transformations to the octonions.
This is accomplished by relating the octonionic \Mobius\ transformations to
the Lorentz group in 10 spacetime dimensions.  The result will be of
particular interest to physicists working with lightlike objects in 10
dimensions.
\endinsert

\Section{INTRODUCTION}

It is well-known that the \Mobius\ transformations generate the conformal
group in the plane, and that they can be identified using stereographic
projection with conformal transformations on the sphere.  As emphasized so
elegantly by Penrose
\Ref{Roger Penrose and Wolfgang Rindler,
{\bf Spinors and Space-Time},
Cambridge University Press, Cambridge, 1984 \& 1986,
and references cited there.} \label\Penrose
in his twistor program, this identification allows the \Mobius\
transformations to be identified with the Lorentz group $\SO(3,1)$ in 4
dimensions, since each Lorentz transformation induces a conformal
transformation on the 2-sphere of null directions.

In this paper we generalize all of this structure in a natural way to the
octonions.  A key piece of the puzzle is the use of the octonionic Lorentz
transformations in 10 dimensions as given by Manogue \& Schray
\Ref{Corinne A. Manogue and J\"org Schray,
{\it Finite Lorentz transformations, automorphisms, and division algebras},
J. Math.\ Phys.\ {\bf 34}, 3746--3767 (1993).}\label\Lorentz
.  We find that, despite the apparent obstacles due to nonassociativity, the
identification of octonionic \Mobius\ transformations with $\SO(9,1)$, is
straightforward.

After much of this work was completed, we discovered the earlier work of
D\"undarer, G\"ursey \& Tze
[\MultiRef{Resit D\"undarer, Feza G\"ursey and Chia-Hsiung Tze,
{\it Self-Duality and Octonionic Analyticity of $S^7$-Valued Antisymmetric
Fields in Eight Dimensions},
Nucl.\ Phys.\ {\bf B266}, 440--450 (1986).}\Label\Dundarer
,\MultiRef{Feza G\"ursey and Chia-Hsiung Tze,
{\bf On the Role of Division, Jordan, and Related Algebras in Particle
Physics}, World Scientific, Singapore, 1996.}]\Label\Tze
, who discuss conformal transformations of $\RR^8$ using similar techniques.
As discussed in more detail below, our treatment differs from theirs in a way
which could have important consequences for the study of lightlike objects in
10 dimensions.  We also point out an error in their treatment of $G_2$, which
invalidates the precise form of the \Mobius\ representation given in
[\Dundarer,\Tze]; this is easily corrected.

In order to keep the article self-contained, we review our basic ingredients
in the first two sections: octonions in Section~2, and complex \Mobius\
transformations in Section~3.  These sections can be safely omitted by the
knowledgeable reader.  In Section~4 we describe the results of Manogue \&
Schray \Lorentz, which show how to resolve the associativity difficulties
inherent in defining {\it finite\/} octonionic Lorentz transformations.  These
same ideas are then used in Section~5 to obtain the new result of this paper,
namely the generalization of \Mobius\ transformations from the complexes to
the octonions.  In Section~6 we discuss the work of D\"undarer, G\"ursey and
Tze, both discussing how our approach differs from theirs and correcting the
aforementioned error.  Finally, we discuss our results in Section~7.

\Section{OCTONIONS}

The {\it octonions\/} $\OO$ are the nonassociative, noncommutative, normed
division algebra over the reals.  In terms of a natural basis, an octonion $a$
can be written
$$a = \sum\limits_{q=1}^8 a^q e_q \Eqno$$
where the coefficients $a^q$ are real, and where the basis vectors satisfy
$e_1=1$ and
$$e_q^2 = -1 \qquad (q=2,...,8) \Eqno$$
We refer to the latter as {\it imaginary basis units}; they anticommute
$$e_q e_r = - e_r e_q \qquad (q\ne r; \, q,r=2,...,8) \Eqno$$
and products of two different imaginary basis units yield a third, i.e.\
$e_q e_r = \pm e_s$ for some $s$.  The full multiplication table is
conveniently encoded in the 7-point projective plane, shown in
\Fig{The representation of the octonionic multiplication table using the
7-point projective plane, where we have used the conventional names
$\{i,j,k,kl,jl,il,l\}$ for $\{e_2,...,e_8\}$.  Each of the 7 oriented lines
gives a quaternionic triple.}
{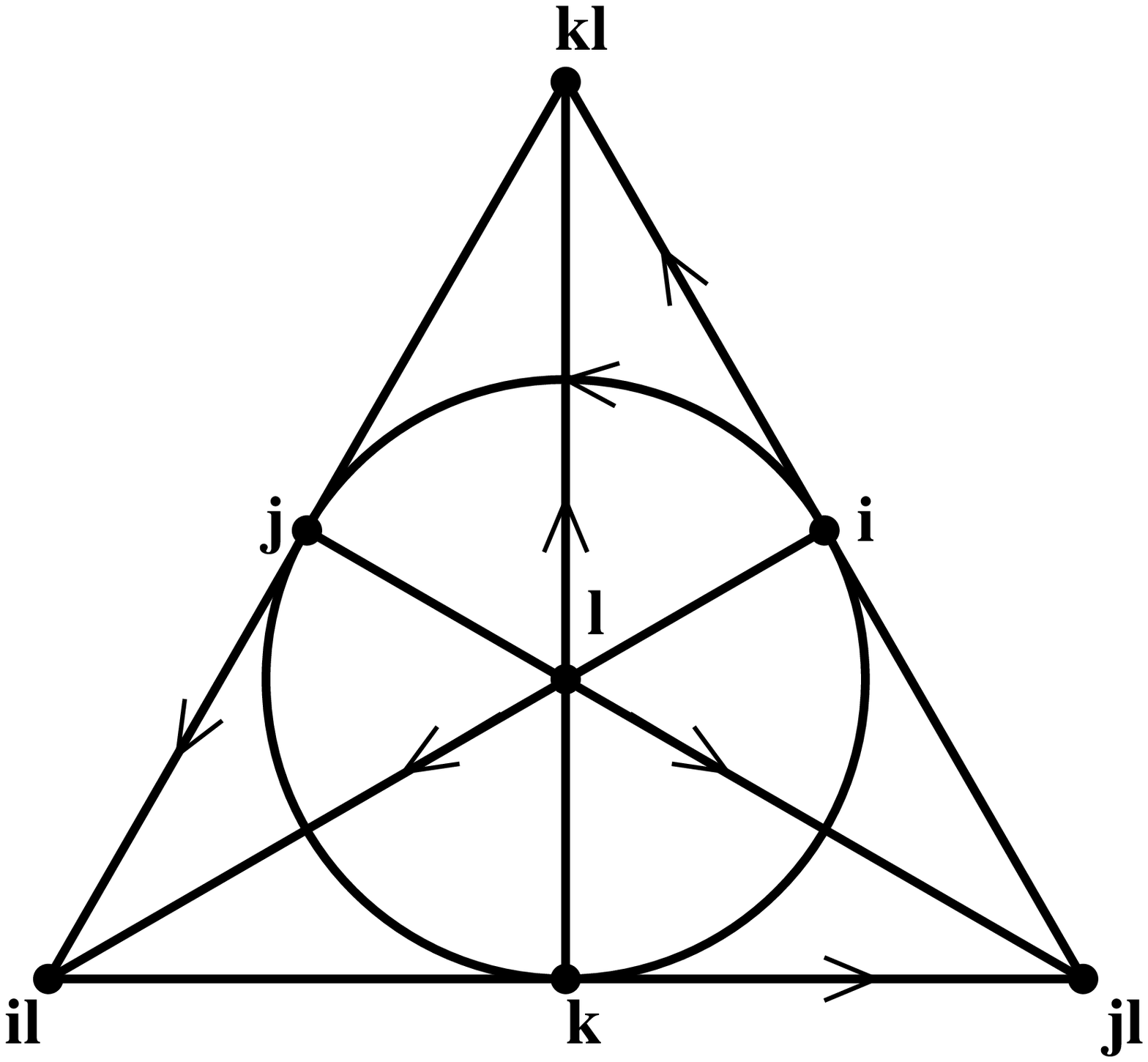}{68 168 543 614}{3in}
{.  The product of any two imaginary units is given by the third unit on the
unique line connecting them, with the sign determined by the relative
orientation.}

The {\it associator\/} of three octonions is
$$[a,b,c] = (ab)c - a(bc) \Eqno$$
which is totally antisymmetric in its arguments and has no real part.
Although the associator does not vanish in general, the octonions do satisfy a
weak form of associativity known as {\it alternativity}, namely
$$[a,b,a]=0 \Eqno$$
The underlying reason for this is that two octonions determine a quaternionic
subalgebra of $\OO$, so that any product containing only two octonionic
directions is associative.

{\it Octonionic conjugation\/} is given by reversing the sign of the imaginary
basis units, so, with $a$ as above,
$$\bar a = a^1 e_1 - \sum\limits_{q=2}^8 a^q e_q \Eqno$$
Conjugation is an antiautomorphism, since it satisfies
$$\bar{ab} = \bar{b} \> \bar{a}$$
The {\it inner product\/} on $\OO$ is the one inherited from $\RR^8$, namely
$$\langle a,b \rangle = \sum_q a^q b^q \Eqno$$
which can be rewritten as
$$\langle a,b \rangle
  = {1\over2} (a \bar{b} + b \bar{a})
  = {1\over2} (\bar{b} a + \bar{a} b)
  \Eqno$$
Finally, the {\it norm\/} of an octonion is just
$$|a| = \sqrt{a \bar{a}} = \sqrt{\langle a,a \rangle} \Eqno$$
which satisfies the defining property of a normed division algebra, namely
$$|ab| = |a| |b| \Eqno$$

\Section{COMPLEX M\"OBIUS TRANSFORMATIONS}

The unit sphere $\SS2\subset\R3$ is related to the Riemann sphere (the complex
plane with a point at infinity added) via {\it stereographic projection\/}
 from the north pole, which takes the point $(x,y,z)$, with $x^2+y^2+z^2=1$, to
the point
$$w = {x+iy \over 1-z} = {1+z \over x-iy} \Eqno$$ \label\SterDef
Under this transformation, the north pole is mapped to
the point at infinity.

As discussed in detail by Penrose and Rindler \Penrose, we can regard $\SS2$
as the set of future (or past) null directions, specifically as the
intersection of the future light cone of the origin in 4-dimensional Minkowski
space with the hypersurface $t=1$.  Other points on a given null ray are
obtained by scaling with $t$, and we can extend stereographic projection to a
map on the entire light cone via
$$w = {x+iy \over t-z} = {t+z \over x-iy} \Eqno$$
with the condition $x^2+y^2+z^2=t^2$.  Penrose and Rindler show how
to obtain this correspondence directly by an orthogonal projection in
Minkowski space, rather than via stereographic projection.

We can further identify $\SS2$ with the complex projective space $\CP1$, the
space of complex lines in $\CC^2$, which is given by
$$\CP1 = \{ [(b,c)] \in \CC^2:
	(b,c) \sim (\xi b,\xi c) \quad \forall\, 0\ne\xi\in\CC \}
  \Eqno$$\label\CPDef
where the square brackets denote equivalence classes under the equivalence
relation $\sim$.  Then each $[(b,c)]\in\CP1$ can be identified with the point
$w$ in the complex plane given by
$$w = {b \over c} \Eqno$$ \label\ProjDef
which is further identified with a point in $\SS2$ via \SterDef; $[(b,0)]$ is
to be identified with the north pole, corresponding to $w=\infty$.
Stereographic projection \SterDef\ can be thought of as a special case of
\ProjDef\ with $b$ or $c$ real.

The \Mobius\ transformations in the complex plane are the complex mappings of
the form~%
\Footnote{An excellent description of these transformations, and their
relation to Lorentz transformations, appears in \S1.2 and \S1.3 of \Penrose.}
$$w \mapsto {\alpha w+\beta \over \gamma w+\delta} \Eqno$$ \label\MobDef
where $\alpha \delta - \beta \gamma \ne 0$.  It is usually assumed without
loss of generality that the complex numbers $\alpha$, $\beta$, $\gamma$,
$\delta$ satisfy
$$\alpha \delta - \beta \gamma = 1 \Eqno$$ \label\Norm
\Mobius\ transformations are the most general analytic transformation of the
{\it Riemann sphere\/} to itself.  Using \ProjDef, we can rewrite \MobDef\ as
$${b \over c} \mapsto {\alpha b + \beta c \over \gamma b + \delta c} \Eqno$$
\label\SpDef
The \Mobius\ transformation \SpDef\ does not depend on the particular
choice of $b$ and $c$ in the equivalence class $[(b,c)]$, which allows us to
view it as acting on $\CP1$.

Now consider a Hermitian $2\times2$ matrix $A$, which we can write as
$$A = \pmatrix{t+z& x+iy\cr \noalign{\smallskip} x-iy& t-z\cr} \Eqno$$
We can identify $A$ with the Minkowski vector $A^\mu=(t,x,y,z)$.
Furthermore, the norm of $A^\mu$ is given by
$$A^\mu A_\mu = -\det A \Eqno$$ \label\DetNorm
where our signature is $(-+++)$.  In particular, $A$ corresponds to a null
direction if and only if $\det A=0$, and in this case we can always write
$$A = v v^\dagger \Eqno$$ \label\VSq
where
$$v = \pmatrix{b\cr c\cr} \Eqno$$ \label\Vdef
is a spinor.  Conversely, the matrix square of any spinor $v$ corresponds to a
null vector.  But the space of null directions is precisely $\SS2$, as can be
seen by simply identifying $\{\xi v\}$ with $w=b/c$, noting that the
overall scale is irrelevant.  (The remaining phase freedom in $b$ and $c$
corresponds to the Hopf fibration.)

Using these various identifications, we can rewrite a \Mobius\ transformation
\MobDef\ as a map on spinors
$$v \mapsto M v \Eqno$$
with $v$ as above and where
$$M = \pmatrix{\alpha& \beta\cr \noalign{\smallskip} \gamma& \delta\cr} \Eqno$$
\label\MatDef
Imposing the condition \Norm, we see that $\det M = 1$, so that $M$ is (the
spinor representation of) a Lorentz transformation.  As could be expected from
\VSq, $M$ acts on vectors $A$ via
$$A \mapsto M A M^\dagger \Eqno$$ \label\MTrans
which preserves the determinant (i.e.\ the norm) of $A$ as required.

We thus see that \Mobius\ transformations are exactly the same as Lorentz
transformations.  Note the key role played by associativity, which allows one
to multiply numerator and denominator of a \Mobius\ transformation by $c$,
thus permitting a reinterpretation as a matrix equation.

\Section{LORENTZ TRANSFORMATIONS}

Manogue \& Schray \Lorentz\ give an explicit representation of a set of
generators for finite Lorentz transformations in 10 spacetime dimensions.
Their results can be summarized as follows.

Let $A$ be a Hermitian $2\times2$ octonionic matrix, so that we can write
$$A = \pmatrix{p& a\cr \noalign{\smallskip} \bar{a}& m\cr} \Eqno$$
where $p$, $m$ are real, and where $a$ is an arbitrary octonion.  Just as in
the complex case, we can associate $A$ with a (real) vector
$A^\mu=(a^0,...,a^9)$ in (10-dimensional) Minkowski space via
$$\eqalign{
  p &= a^0 + a^9 \cr
  m &= a^0 - a^9 \cr
  a &= \sum\limits_{q=1}^8 a^q e_q
  }\Eqno$$
As in the complex case, the norm of $A^\mu$ is given by \DetNorm; the
determinant is well-defined since $A$ is Hermitian.  If $A$ is null we still
have \VSq\ and \Vdef; the freedom in choosing $b$ and $c$ again corresponds to
a real scale and the Hopf fibration (of $\SS{15}$ in this case).

A {\it Lorentz transformation\/} $M$ acts on $A$ via \MTrans, and leaves the
determinant invariant, thus preserving the norm of $A^\mu$.  The first
requirement on $M$ is that this be well-defined, i.e.\ that
$$(MA) M^\dagger = M (AM^\dagger) =: MAM^\dagger \Eqno$$
In particular, this means that $MAM^\dagger$ is indeed Hermitian.  Manogue \&
Schray note that for this to be the case, either $M$ must be {\it complex},
i.e.\ the components of $M$ lie in a complex subspace of $\OO$, or the columns
of the imaginary part of $M$ must be real multiples of each other.  In either
case,
$$\det(MAM^\dagger) = \det(MM^\dagger) \det A \Eqno$$
so that a further condition for $M$ to be a Lorentz transformation is
$$\det(MM^\dagger) = 1 \Eqno$$\label\DD

It is well-known, however, that not all Lorentz transformations in 10
dimensions can be written in the form \MTrans.  Because of the lack of
associativity, not all Lorentz transformations can be achieved with a single
matrix $M$.  Manogue \& Schray \Lorentz\ show, however, that {\it nested\/}
transformations are sufficient, e.g.\
$$A \mapsto M_n( ... (M_1 A M_1^\dagger) ... )M_n^\dagger \Eqno$$
The generating set they give requires at most $n=2$, and that only for the
transverse rotations which interchange imaginary octonionic units.  As they
show, these rotations can be generated by two {\it flips}, each of which is a
pure imaginary multiple of the identity matrix.

Turning to the action of the Lorentz group on spinors $v$, there is another
associativity problem.  The generators $M$ given by Manogue \& Schray
\Lorentz\ satisfy a {\it compatibility\/} condition between the spinor and
vector representations, namely
$$(M v) (M v)^\dagger = M (v v^\dagger) M^\dagger \Eqno$$ \label\Compat
It can be shown that the necessary and sufficient conditions for $M$ to be
compatible are that its components all lie in a single complex subspace of the
octonions, and that
\Footnote{This result was checked using {\sl Mathematica}.}
$$\det M \in \RR \Eqno$$ \label\DetEq
Note that the determinant is well-defined here because $M$ is complex.
\Footnote{It is intriguing to compare \DD\ with Dieudonn\'e's prescription
\Ref{Emil Artin, {\bf Geometric Algebra},
John Wiley \& Sons, New York, 1957 \& 1988.}
for the determinant $\Det(M)$ of a quaternionic matrix $M$, which reduces in
the $2\times2$ case to
(compare
\Ref{E. Study, Acta Math.\ {\bf 42}, 1 (1920);
\hfill\break
Freeman J. Dyson, {\it Quaternion Determinants},
Helv.\ Phys.\ Acta {\bf 45}, 289--302 (1972).})
$$\Det(M)=\sqrt{\det(MM^\dagger)}$$
where of course $\det(MM^\dagger)$ denotes the ordinary (complex) determinant
of $MM^\dagger$, which is a positive real number.  While the Dieudonn\'e
determinant has some nice properties, including the fact that
$$\Det(MN) = \Det(M) \, \Det(N)$$
we choose to work with the ordinary determinant in large part due to the
compatibility condition \DetEq, which can not be expressed using the
Dieudonn\'e determinant alone, which is always real and positive.}

Thus, the (finite) Lorentz transformations are generated by all $2\times2$
complex matrices which have determinant $\pm1$.  (There is of course no
requirement that these matrices all lie in the {\it same\/} complex subspace.)
It is interesting to compare this with the complex and quaternionic cases,
where compatibility is automatic.  In the complex case, multiplication of $M$
by an arbitrary phase $e^{i\theta}$ does not change the vector transformation,
but does affect the spinor transformation.  The determinant condition \DetEq\
eliminates all of these transformations except those for which
$\theta=n\pi/2$.  The half-integer multiples of $\pi$ can be eliminated by
requiring the determinant to be $+1$, leaving only the expected 2-to-1 mapping
corresponding to an arbitrary overall sign.

The quaternionic case is more subtle: Multiplication by $e^{i\theta}$ now
corresponds to a rotation in the $jk$-plane, and therefore must be included as
a Lorentz transformation even though its determinant is not real.  It is
interesting to note that the work of Manogue \& Schray shows how to write
such transformations as a product of two flips, each of which has determinant
$-1$; the determinant of a product fails to be the product of the determinants
in this case.  Thus, \DetEq\ can indeed be used to define the generators of
(finite) Lorentz transformations in this case.

Returning to the octonionic case, all of Manogue \& Schray's generators are
complex and either have determinant $+1$ or are constructed from two nested
transformations, each of which has determinant $-1$.  Thus, the Lorentz group
could be defined for each of the division algebras as being generated by such
matrices.  Furthermore, it is straightforward to write those of Manogue \&
Schray's generators with determinant $+1$ as the product of two transformations
with determinant $-1$, and it is interesting that, in the octonionic case, the
flips themselves can be so generated, even though they have determinant $-1$.

We can therefore define the Lorentz group in each case to be those
transformations generated by {\bf two} complex matrices of determinant $-1$,
which suitably generalizes the more traditional definition in terms of
matrices of determinant $+1$.  It is only in this nested sense that
$\SL(2,\OO)$ consists of ``all matrices of determinant $+1$.''
\Footnote{There is a notational hazard here: $\SL(2,\OO)$ could refer to either
of two quite different objects, an ambiguity which does not arise over the
other division algebras.  The first possibility is the matrix algebra of the
$2\times2$ matrices just discussed, which is not associative, and hence not a
group.  The second possibility is the action of these matrices on either
spinors ($2$-component octonionic columns) or vectors ($2\times2$ octonionic
Hermitian matrices).  This {\it is\/} a group; the group operation is
composition, which is associative.  The main result of \Lorentz\ is that this
group is (isomorphic to) the double cover of $\SO(9,1)$, and hence is
isomorphic to $\Spin(9,1)$.  It is attractive to write this isomorphism as
$\SL(2,\OO)\approx\Spin(9,1)$, in which $\SL(2,\OO)$ refers to the second
possibility; we feel that the current paper further supports this usage.
However, this notational ambiguity does {\it not\/} affect the statement made
in the main text above.}

\Section{OCTONIONIC M\"OBIUS TRANSFORMATIONS}

Putting this all together, we will invert the usual derivation that Lorentz
transformations are the same as \Mobius\ transformations.  Rather, we will
{\it define\/} octonionic \Mobius\ transformations in terms of the Lorentz
transformations of Manogue \& Schray, and then show that these
transformations can be rewritten in the form \SpDef.

Thus, given an octonion $w$, define (generators of) \Mobius\ transformations
via \MobDef, which we rewrite as
$$f_M(w) = (\alpha w+\beta) (\gamma w+\delta)^{-1} \Eqno$$\label\WTrans
and where the matrix of coefficients $M$ defined by \MatDef\ is now not
only octonionic, but is further required to be one of Manogue \& Schray's
compatible generators of the Lorentz group.

We would like to be able to construct more general \Mobius\ transformations by
nesting.  However, it is not at all obvious that iterating \WTrans\ leads to a
(suitably nested) transformation of the same type.  We would really like to be
able to use (an octonionic version of) \SpDef\ to define \Mobius\
transformations, as this would make it apparent that iterating \Mobius\
transformations corresponds directly to nesting Lorentz transformations.  As
previously noted, this requires \SpDef\ to be independent of the particular
choice of $b$ and $c$.  Remarkably, the octonionic generalization of \SpDef\
does have this property, as we now show; this is our main result.

Suppose that
$$w = b c^{-1} \Eqno$$\label\ProjDefO
where now $b,c\in\OO$, and construct the spinor $v$ as in \Vdef.  Letting
$$v_0 = \pmatrix{w\cr 1\cr} \Eqno$$
we have
$$v = v_0 c \Eqno$$
and
$$v v^\dagger = |c|^2 v_0 v_0^\dagger \Eqno$$
since only two octonionic directions are involved.

We now write
$$V = Mv = \pmatrix{B\cr C\cr} = \pmatrix{BC^{-1}\cr 1\cr} C \Eqno$$
leading to
$$V V^\dagger
  = \pmatrix{|B|^2& B\bar{C}\cr \noalign{\smallskip} C\bar{B}& |C|^2\cr}
  = |C|^2 \pmatrix{{|B|^2\over|C|^2}& {\scriptstyle BC^{-1}}\cr 
		\noalign{\smallskip} {\scriptstyle\bar{BC^{-1}}}& 1\cr}
  \Eqno$$
and similar relations for $V_0=Mv_0$.  Compatibility now leads to
$$\eqalign{
  V V^\dagger
  &= (Mv) (Mv)^\dagger \cr
  &= M (v v^\dagger) M^\dagger
   = |c|^2 M (v_0 v_0^\dagger) M^\dagger \cr
  &= |c|^2 (Mv_0) (Mv_0)^\dagger
   = |c|^2 V_0 V_0^\dagger
  }\Eqno$$ \label\MVeq
Comparing the offdiagonal entries of \MVeq, we obtain
$$|C|^2 B C^{-1} = |c|^2 |C_0|^2 B_0 C_0^{-1} \Eqno$$
But direct computation shows that
$$|C|^2 = |\gamma b + \delta c|^2
	= |\gamma w + \delta|^2 |c|^2 = |C_0|^2 |c|^2
  \Eqno$$
provided
$$\Big\langle [b,c,\gamma] , \delta \Big\rangle = 0\Eqno$$
which holds for compatible $M$ since $\gamma$ and $\delta$ lie in the same
complex subspace of $\OO$.  Finally, by construction we have
$$f_M(w) = B_0 C_0^{-1} \Eqno$$
and putting this all together results in
$$B C^{-1} = f_M(w) \Eqno$$
or equivalently
$$f_M(w)
  = (\alpha w+\beta) (\gamma w+\delta)^{-1}
  = (\alpha b+\beta c) (\gamma b+\delta c)^{-1}
  \Eqno$$
This is the desired result, since $b$ and $c$ were arbitrary (satisfying
\ProjDefO).

\Section{PREVIOUS WORK}

D\"undarer, G\"ursey, \& Tze [\Dundarer,\Tze] give a \Mobius\ representation
of conformal transformations in $\RR^8$, which relies on a decomposition of
the form
$$G_2 \subset \Spin(7) \subset \Spin(8) \subset \Spin(9,1) \Eqno$$
They thus reduce an arbitrary conformal transformation to a composition of the
form [\Dundarer,\Tze]
\Footnote{The order of $K$ and $L$ can presumably be reversed, although this
will change the transformation.}
$$x \mapsto
	(UV)^{-1} \left\{ V \left( U \left[ K \left( L
	\left( {\lambda\over x-A} + \bar{C} \right)^{-1}
	\bar{L} \right) K \right] U^{-1} \right) V^{-1} \right\} (UV)
  \Eqno$$ \label\GT
where the parameters $\lambda\in\RR$, $A,C,K,L\in\OO$ with $|K|=1=|L|$
correspond to a dilation, a translation, a special conformal transformation, a
rotation in $\Spin(8)/\Spin(7)$, and a rotation in $\Spin(7)/G_2$,
respectively.  They claim the remaining parameters $U,V\in\OO$ correspond to a
$G_2$ transformation, which they give in three forms (page 229 of [\Tze]),
related by triality
$$\eqalignno{
y &\mapsto
  (\bar{ab}) \left[ b \Bigl( ay\bar{a} \Bigr) \bar{b} \right] (ab)
  \Eqalignno\cr\label\Gtwo
y &\mapsto
  (\bar{ab}) \left[ b \left( aya^2 \right) b^2 \right] (\bar{ab})^2
  \Eqalignno\cr\label\SpinorI
y &\mapsto
  (ab)^2 \left[ \bar{b}^2 \left( \bar{a}^2y\bar{a} \right) \bar{b} \right] (ab)
  \Eqalignno\label\SpinorII
  }$$
where $|a|=1=|b|$.
\Footnote{The expression \GT\ uses the first form \Gtwo\ with $a={U\over|U|}$,
$b={V\over|V|}$.}
G\"ursey \& Tze [\Tze] note correctly that {\it if\/} both of $a$ and $b$
admit power series expansions around 1, {\it then\/} the infinitesimal form of
each of \Gtwo--\SpinorII\ agrees with the standard form of $G_2$ as the
derivation algebra of $\OO$
\Ref{Richard D. Schafer,
{\bf An Introduction to Nonassociative Algebras}, Academic Press, New York,
1966 \& Dover, Mineola NY, 1995.}%
.  However, as we now show, the expressions \Gtwo--\SpinorII\ are {\it not\/}
automorphisms for all values of $a$ and $b$; the assumed power series
expansions are not valid.

{\let\ell=e The two octonions $a$ and $b$ span a quaternionic subspace
$\HH\subset\OO$.  We can therefore view the octonions as arising from $\HH$
via the Cayley-Dickson process, so that
$$\OO = \HH \oplus \HH\ell \Eqno$$
where $\ell$ is any pure imaginary octonionic unit orthogonal to $\HH$.  We
can thus write any octonion $x\in\OO$ uniquely as
$$x = x_1 + x_2 e \Eqno$$
with $x_i\in\HH$, and the multiplication of two such octonions $x$, $y$ can be
written as
$$x y
  = (x_1 + x_2 e) (y_1 + y_2 e)
  = (x_1 y_1 - \bar{y}_2 x_2) + (y_2 x_1 + x_2 \bar{y}_1) e
  \Eqno$$
}%
Rewriting \Gtwo--\SpinorII\ in this way, it is lengthy but straightforward to
show that \Gtwo--\SpinorII\ are automorphisms if and only if
\Footnote{Interestingly, this is not just an associativity issue:
\SpinorI--\SpinorII\ are not automorphisms of $\HH$ unless this condition
holds, although \Gtwo\ is.}
$$a b a b a = b a^3 b \Eqno$$\label\TD
Furthermore, \TD\ is identically satisfied to second order, thus confirming
second-order agreement with the derivation algebra as claimed in [\Tze].  If
$a$, $b$ lie in a {\it complex\/} subspace of $\OO$, \TD\ is trivially
satisfied but, using alternativity, each of \Gtwo--\SpinorII\ reduces to the
identity.  However, except for this special case, it turns out that $a$ must
be purely imaginary.  Thus, in no case does an automorphism of the form
\Gtwo--\SpinorII\ admit the assumed power series expansion!
\Footnote{The first form \Gtwo\ is also an automorphism if $ababa=-ba^3b$, in
which case {\it both\/} $a$ and $b$ are bounded away from 1.}
We therefore find it remarkable that not only do automorphisms of the form
\Gtwo\ exist at all, but enough such automorphisms exist to generate all of
$G_2$ (by iteration).

However, it is not necessary for the construction of D\"undarer, G\"ursey, \&
Tze that \Gtwo\ be an automorphism for all values of $a$ and $b$.  Rather, it
would be enough if every $G_2$ transformation could be written in this form.
However, this also fails: All of these automorphisms fix precisely 1
octonionic direction (except for special values of the parameters), but there
are automorphisms which leave entire quaternionic subspaces invariant.  We
conclude that the decomposition of G\"ursey \& Tze must be modified so as to
include (at least) one additional $G_2$ transformation.

We now compare this treatment of $G_2$ with that of Manogue \& Schray
\Lorentz, who show how to generate the group $\Spin(7)$ by nesting either
conjugation, left multiplication, or right multiplication.  Since elements of
$G_2\subset\Spin(7)$ can be generated by nesting opposite $\Spin(7)$ rotations
in two planes which ``point'' towards the same octonionic direction, this
yields a nested representation of $G_2$.  Rewriting these in a form similar to
\hbox{\Gtwo--\SpinorII}, we obtain
$$\eqalignno{
y &\mapsto
  (d\ell) \left( (c\ell) \left[ d
	\left( c y \bar{c} \right)
		\bar{d} \right] (\bar{c\ell}) \right) (\bar{d\ell})
  \Eqalignno\cr\label\CAMI
y &\mapsto
  (d\ell) \Bigl( (c\ell) \Bigl[ d
	\left( c y \right)
		\Bigr] \Bigr)
  \Eqalignno\cr
y &\mapsto
  \left( \left[ 
	\left( y \bar{c} \right)
		\bar{d} \right] (\bar{c\ell}) \right) (\bar{d\ell})
  \Eqalignno\label\CAMII
  }$$
where $c$, $d$ are pure imaginary unit octonions and $\ell$ is any imaginary
unit octonion orthogonal to the quaternionic subspace spanned by $c$, $d$.  If
in addition $c$ and $d$ are orthogonal, then \hbox{\CAMI--\CAMII} agree
exactly with \Gtwo--\SpinorII\ under the identification $c=ae$, $d=be$, with
$e$ orthogonal to $\HH$ as before, and where $\ell$ is any (normalized) linear
combination of $a$ and $b$.  In general, however, \Gtwo--\SpinorII\ and
\CAMI--\CAMII\ give different ``bases'' for $G_2$, since the latter fix a
quaternionic subspace and the former do not.  In any case, all \Mobius\
representations can also be obtained by iterating \GT.

\Section{DISCUSSION}

We have shown that the finite octonionic Lorentz transformations in 10
dimensions as given by Manogue \& Schray~\Lorentz\ can be used to define
octonionic \Mobius\ transformations, thus recovering (and correcting) the
earlier results of D\"undarer, G\"ursey, \& Tze [\Dundarer,\Tze].  However,
our approach differs significantly from theirs, as theirs corresponds to using
\MobDef, while ours uses~\SpDef.  We have thus shown that octonionic \Mobius\
transformations extend to the octonionic projective space $\OP1$, defined by
\Footnote{A related definition in terms of $2\times2$ octonionic Hermitian
matrices (the ``square'' of the form given here) was given by Harvey (page 123
of
\Ref{F. Reese Harvey,
{\bf Spinors and Calibrations}, Academic Press, Boston, 1990.}%
).  However, there is a minor error in his discussion of the equivalence
relation, which we have corrected.}
$$\OP1 = \{ [(b,c)] \in \OO^2:
	(b,c) \sim \left((bc^{-1})\xi,\xi\right)
		\quad \forall\, 0\ne\xi\in\OO \}
  \Eqno$$
We believe that this may be the key result needed to generalize 4-dimensional
twistor theory to 10 dimensions.  Much recent research in superstrings,
supergravity, and M-theory has emphasized the importance of lightlike objects
in 10 dimensions.  An appropriate octonionic generalization of twistor theory
to 10 dimensions might allow powerful twistor techniques to be applied to
these other theories.

A key role in our argument is the use of two fundamental properties of the
octonionic Lorentz transformations in \Lorentz, namely {\it nesting\/} and
{\it compatibility}.  Our results here support our view that these are
essential features of any computation involving octonions.  Otherwise,
repeated transformations of the form \MobDef\ are not equivalent to those of
the form \SpDef, due to the lack of associativity.

Finally, by restricting to a quaternionic subspace of the octonions, we also
obtain a corresponding relationship between quaternionic \Mobius\
transformations and the Lorentz group $\SO(5,1)$ in 6 dimensions.  Since there
are no associativity problems, this could of course have been obtained by
straightforward generalization of the complex case.

\bigskip
\References

\bye